\documentclass[conference]{IEEEtran}
\IEEEoverridecommandlockouts
\usepackage{cite}
\usepackage{amsmath,amssymb,amsfonts}
\usepackage{algorithmic}
\usepackage{graphicx}
\usepackage{textcomp}
\usepackage{xcolor}
\def\BibTeX{{\rm B\kern-.05em{\sc i\kern-.025em b}\kern-.08em
    T\kern-.1667em\lower.7ex\hbox{E}\kern-.125emX}}
\begin{document}

\title{\LARGE Transformer-Based Hybrid Beamforming with Reconfigurable Pixel Antenna for HAPS Communications
}

\author{Ruiqi Wang, Ziwei Wan, Keke Ying and Zhen Gao

}

\maketitle

\begin{abstract}
This paper proposes a Transformer-based hybrid beamforming framework for reconfigurable  pixel antenna (RPA)-equipped massive multiple-input multiple-output (MIMO) in high-altitude platform station (HAPS) communications. The proposed pattern reconfigurable hybrid beamforming network (PR-HBFNet) comprises two key components: 1) a pattern reconfigurable network that leverages a Transformer encoder to determine the radiation pattern for each antenna element, and 2) a hybrid beamforming network that employs model-driven residual learning to compute analog and digital precoders over SVD-based initializations. Simulation results demonstrate that the proposed PR-HBFNet closely approaches the spectral efficiency of a greedy benchmark while significantly reducing computational complexity.
\end{abstract}


\section{Introduction}

High-altitude platform stations (HAPS), operating in the stratosphere at altitudes of 17--22 km, have emerged as a promising component of the integrated space-air-ground network envisioned for sixth-generation (6G) wireless systems \cite{kurt2021vision, arum2020review}. Owing to their quasi-stationary deployment, wide coverage footprint, and strong probability of line-of-sight (LoS) connectivity, HAPS can effectively bridge the coverage and capacity gap between terrestrial infrastructure and low-earth-orbit (LEO) satellite systems \cite{arum2020review}. With the International Telecommunication Union (ITU) allocating dedicated spectrum resources for HAPS communications in both sub-6 GHz and millimeter-wave (mmWave) bands \cite{itu2019haps}, the design of high-throughput and energy-efficient air-to-ground transmission techniques has become an important research priority. 

To meet the stringent data-rate requirements of HAPS links while compensating for severe free-space path loss, massive multiple-input multiple-output (MIMO) has been recognized as a key enabling technology \cite{wang2026transformer}. However, fully digital beamforming is often impractical for HAPS payloads because it requires a dedicated radio-frequency (RF) chain per antenna, resulting in excessive hardware cost and power consumption. Hybrid beamforming (HBF), which distributes precoding across a low-dimensional digital precoder and a high-dimensional analog beamforming network, provides a more practical solution by significantly reducing the number of RF chains while retaining most of the beamforming gain \cite{el2014spatially}. Nevertheless, conventional HBF architectures are typically built upon fixed-pattern antenna elements whose radiation characteristics remain unchanged during operation. In HAPS air-to-ground channels, where propagation is usually dominated by LoS or near-LoS components with limited angular spread, such static radiation patterns restrict the ability of the antenna array to further concentrate energy toward the dominant spatial directions, thereby leaving additional array and channel gains underexploited \cite{ying2024reconfigurable}. 

RPAs provide a promising means to overcome this limitation. By integrating electronically controlled switches into the antenna structure, each element can dynamically select from a predefined set of radiation patterns, thereby introducing an additional degree of freedom in the electromagnetic (EM) domain beyond conventional amplitude-and-phase precoding \cite{zhang2022highly, ying2024reconfigurable}. For HAPS systems serving ground users through sparse dominant propagation paths, this EM-domain flexibility enables the array to better align its intrinsic radiation behavior with the channel structure, thus improving effective channel gain and spectral efficiency. Consequently, RPA-assisted HBF motivates a joint optimization framework spanning three coupled domains: pattern selection, analog beamforming, and digital precoding. 

However, the joint design of RPA states and hybrid precoders is highly challenging. Since pattern selection is discrete while analog and digital beamforming variables are continuous, the resulting problem is naturally formulated as a mixed-integer nonlinear programming (MINLP) problem, whose search space grows exponentially with the number of antennas and available radiation modes \cite{shen2017successive}. Although existing iterative algorithms, such as greedy pattern selection combined with conventional HBF optimization, offer tractable approximations, they still rely on repeated matrix operations and complex computations during the optimization process, which leads to substantial computational complexity and inference latency \cite{ying2024reconfigurable}. Such latency is difficult to accommodate in practical HAPS systems with stringent real-time scheduling requirements. 

To address the above challenges, this paper proposes a Transformer-based end-to-end framework to solve the joint pattern--analog--digital beamforming problem \cite{vaswani2017attention, wang2023transformer6g}. Specifically, we develop a cascaded network integrating Transformer encoders with model-driven residual learning to jointly optimize radiation pattern selection and hybrid beamforming. In the proposed design, the sub-connected analog architecture is preserved explicitly and the digital precoders are normalized according to the total transmit-power constraint over all subcarriers. Simulation results verify that the proposed PR-HBFNet closely matches a greedy benchmark in spectral efficiency while significantly reducing computational complexity.

\section{System Model and Problem Statement}

\subsection{Antenna Architecture}

\begin{figure}[tbp]
\centerline{\includegraphics[width=0.4\textwidth]{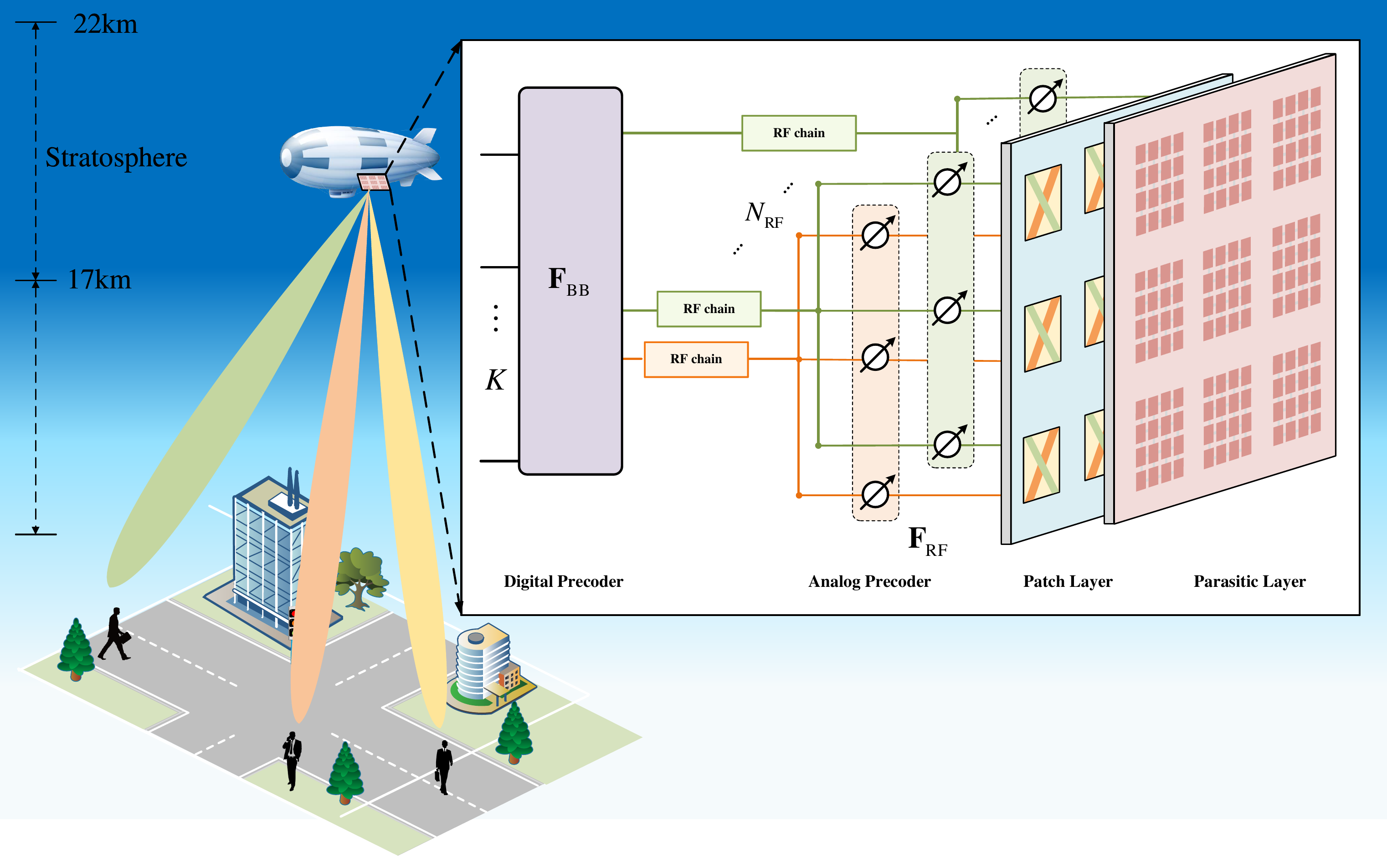}}
\vspace{-3mm}
\caption{HAPS communication system using massive MIMO with reconfigurable antennas.}
\label{fig_haps}
\end{figure}

Consider a stratospheric HAPS as illustrated in Fig.~\ref{fig_haps}, located at a stable altitude serving $K$ single-antenna ground user equipments (UEs) in the downlink. The HAPS is equipped with a uniform planar array (UPA) consisting of $N_t$ RPA elements. To comply with the stringent payload and power constraints inherent to stratospheric platforms, a sub-connected hybrid beamforming architecture is employed. Specifically, the $N_t$ antennas are partitioned into $N_{\text{RF}}$ disjoint subarrays, where each subarray contains $N_s = N_t / N_{\text{RF}}$ elements and is driven by a single RF chain. We assume that the number of RF chains satisfies $K \leq N_{\text{RF}} \ll N_t$. 

Unlike conventional static arrays, each element in the UPA adopts a reconfigurable parasitic pixel structure. This architecture features an active central radiating patch surrounded by an overlying parasitic layer of electrically small conductive pixel units. These units are interconnected via high-speed, low-loss PIN diode switches \cite{zhang2022highly}. By configuring the on/off states of these diodes through baseband digital control signals, the surface current distribution across the element is reshaped within microseconds. This EM-domain tuning enables each $m$-th antenna element to rapidly switch among $M$ pre-optimized candidate radiation patterns. 

Let $\mathcal{M} = \{1, 2, \ldots, M\}$ denote the predefined codebook of available discrete radiation patterns. The specific radiation pattern mode selected by the $m$-th antenna element is denoted as $c_m \in \mathcal{M}$. The global EM configuration of the entire HAPS massive MIMO array is completely characterized by the discrete pattern selection vector $\mathbf{c} = [c_1, c_2, \ldots, c_{N_t}]^T \in \mathcal{M}^{N_t}$. 

\subsection{Channel Model}

The air-to-ground communication link operates over an Orthogonal Frequency-Division Multiplexing (OFDM) system with $N_c$ subcarriers and a total bandwidth of $B_w$. Given the high-altitude vantage point of the HAPS, the propagation environment exhibits sparse angular spreads and is dominated by a strong LoS path, alongside a limited number of non-line-of-sight (NLoS) scattering clusters \cite{kurt2021vision}. To capture these spatial-temporal characteristics, a wideband 3D geometry-based stochastic model (GBSM) experiencing Rician fading is adopted \cite{lian2018gbsm}. 

Assuming the operating bandwidth $B_w$ is significantly smaller than the central carrier frequency $f_c$, the spatial phase shifts between antenna elements are accurately modeled using the central wavelength $\lambda_c = c/f_c$. Leveraging this narrow-band array assumption, we define a frequency-invariant spatial-electromagnetic steering vector $\mathbf{a}(\theta, \phi, \mathbf{c}) \in \mathbb{C}^{N_t \times 1}$ whose $m$-th element is expressed as

\begin{equation}
[\mathbf{a}(\theta, \phi, \mathbf{c})]_m = G_{c_m}(\theta, \phi) e^{-j\frac{2\pi}{\lambda_c}\mathbf{k}^T \mathbf{p}_m},
\end{equation}

where $G_{c_m}(\theta, \phi)$ represents the complex radiation pattern gain of the $m$-th antenna under its active discrete mode $c_m \in \mathcal{M}$, evaluated at the elevation angle $\theta$ and azimuth angle $\phi$. Additionally, $\mathbf{k}$ is the directional unit wave vector, and $\mathbf{p}_m$ denotes the physical 3D coordinate of the $m$-th transmit antenna. 

By utilizing this steering vector, the wideband frequency-domain channel $\mathbf{h}_{u,g}(\mathbf{c}) \in \mathbb{C}^{N_t \times 1}$ between the HAPS array configured in state $\mathbf{c}$ and the $u$-th user on subcarrier $g$ is formulated as the superposition of the deterministic LoS component (index $0$) and the scattered NLoS components (indices $1, \dots, L$):
\begin{equation}
\begin{aligned}
\mathbf{h}_{u,g}(\mathbf{c}) = & \sqrt{\frac{K_R}{K_R+1}} \, \beta_{0,u,g} \, \mathbf{a}(\theta_{0,u}, \phi_{0,u}, \mathbf{c})+ \\
& \sqrt{\frac{1}{K_R+1}} \sum_{l=1}^{L} \beta_{l,u,g} \, \mathbf{a}(\theta_{l,u}, \phi_{l,u}, \mathbf{c}),
\end{aligned}
\end{equation}
where $K_R$ is the Rician factor defining the power ratio between the LoS and NLoS paths. The term $\beta_{l,u,g} = \alpha_{l,u} e^{-j2\pi\tau_{l,u}f_g}$ encapsulates the complex path gain $\alpha_{l,u}$ and the frequency-selective phase shift induced by the propagation delay $\tau_{l,u}$ on subcarrier frequency $f_g$. By stacking the channels of all $K$ users, the aggregate downlink channel matrix under the discrete pattern $\mathbf{c}$ on subcarrier $g$ is defined as $\mathbf{H}_g(\mathbf{c}) = [\mathbf{h}_{1,g}(\mathbf{c}), \dots, \mathbf{h}_{K,g}(\mathbf{c})]^H \in \mathbb{C}^{K \times N_t}$. 

For subsequent learning, we further collect the channel responses associated with all candidate radiation patterns into an electromagnetic channel state information tensor $\mathbf{\hat{H}} \in \mathbb{C}^{N_c \times K \times N_t \times M}$, where the last dimension corresponds to the candidate pattern index. Once the pattern vector $\mathbf{c}$ is determined, the equivalent channel used for beamforming is obtained by selecting, for each antenna element, the channel coefficient associated with its active radiation mode.

\subsection{Problem Formulation}
For downlink transmission, the data symbols intended for the $K$ users on subcarrier $g$, denoted by $\mathbf{s}_g \in \mathbb{C}^{K \times 1}$ where $\mathbb{E}[\mathbf{s}_g\mathbf{s}_g^H] = \mathbf{I}_K$, are first precoded in the digital domain using the baseband precoder $\mathbf{F}_{\text{BB},g} = [\mathbf{f}_{1,g}, \dots, \mathbf{f}_{K,g}] \in \mathbb{C}^{N_{\text{RF}} \times K}$. The signal is subsequently mapped to the antenna elements via the analog RF precoder $\mathbf{F}_{\text{RF}} \in \mathbb{C}^{N_t \times N_{\text{RF}}}$. Due to the sub-connected architecture using passive phase shifters, $\mathbf{F}_{\text{RF}}$ is shared across all subcarriers and has a block-diagonal structure with constant-modulus non-zero entries \cite{el2014spatially}. 

The received baseband signal for user $u$ on subcarrier $g$ is subject to additive white Gaussian noise (AWGN) $n_{u,g} \sim \mathcal{CN}(0, \sigma^2)$. The received signal $y_{u,g}$ is modeled as:

\begin{equation}
\begin{aligned}
y_{u,g} = & \mathbf{h}_{u,g}^H(\mathbf{c}) \mathbf{F}_{\text{RF}} \mathbf{f}_{u,g} s_{u,g} \\
& + \sum_{k \neq u} \mathbf{h}_{u,g}^H(\mathbf{c}) \mathbf{F}_{\text{RF}} \mathbf{f}_{k,g} s_{k,g} + n_{u,g}.
\end{aligned}
\end{equation}

Consequently, the signal-to-interference-plus-noise ratio (SINR) evaluates to:
\begin{equation}
\gamma_{u,g} = \frac{\left|\mathbf{h}_{u,g}^H(\mathbf{c}) \mathbf{F}_{\text{RF}} \mathbf{f}_{u,g}\right|^2}{\sum_{k \neq u} \left|\mathbf{h}_{u,g}^H(\mathbf{c}) \mathbf{F}_{\text{RF}} \mathbf{f}_{k,g}\right|^2 + \sigma^2}.
\end{equation}

The objective of the HAPS communication system is to jointly optimize the discrete pattern selection vector $\mathbf{c}$, the analog precoder $\mathbf{F}_{\text{RF}}$, and the subcarrier-specific digital precoders $\{\mathbf{F}_{\text{BB},g}\}$ to maximize the average downlink sum spectral efficiency. The problem is formulated as
\begin{equation}
\begin{aligned}
\max_{\mathbf{c},\mathbf{F}_{\text{RF}},\{\mathbf{F}_{\text{BB},g}\}} \quad & \frac{1}{N_c} \sum_{g=1}^{N_c} \sum_{u=1}^{K} \log_2\left(1 + \gamma_{u,g}\right) \\
\text{s.t.} \quad & c_m \in \mathcal{M}, \quad \forall m \in \{1, \dots, N_t\}, \\
& \mathbf{F}_{\text{RF}} \in \mathcal{A}_{\mathrm{sub}}, \\
& \frac{1}{N_c}\sum_{g=1}^{N_c} \left\| \mathbf{F}_{\text{RF}} \mathbf{F}_{\text{BB},g} \right\|_F^2 \leq P_t,
\end{aligned}
\label{eq:opt_prob}
\end{equation}
where $\mathcal{A}_{\mathrm{sub}}$ denotes the feasible set of block-diagonal analog precoders under the sub-connected architecture, and each non-zero entry of $\mathbf{F}_{\text{RF}}$ has constant modulus. 

This formulation yields a highly complex MINLP problem. The discrete combinatorial search space for $\mathbf{c}$ scales exponentially as $M^{N_t}$, and it is coupled with the non-convex hardware constraint of the analog beamformer. Resolving this optimization problem in real time motivates the deep learning approach proposed in the following section. 

\section{Proposed Pattern Reconfigurable Hybrid Beamforming Network}
\begin{figure}[tbp]
\centerline{\includegraphics[width=0.5\textwidth]{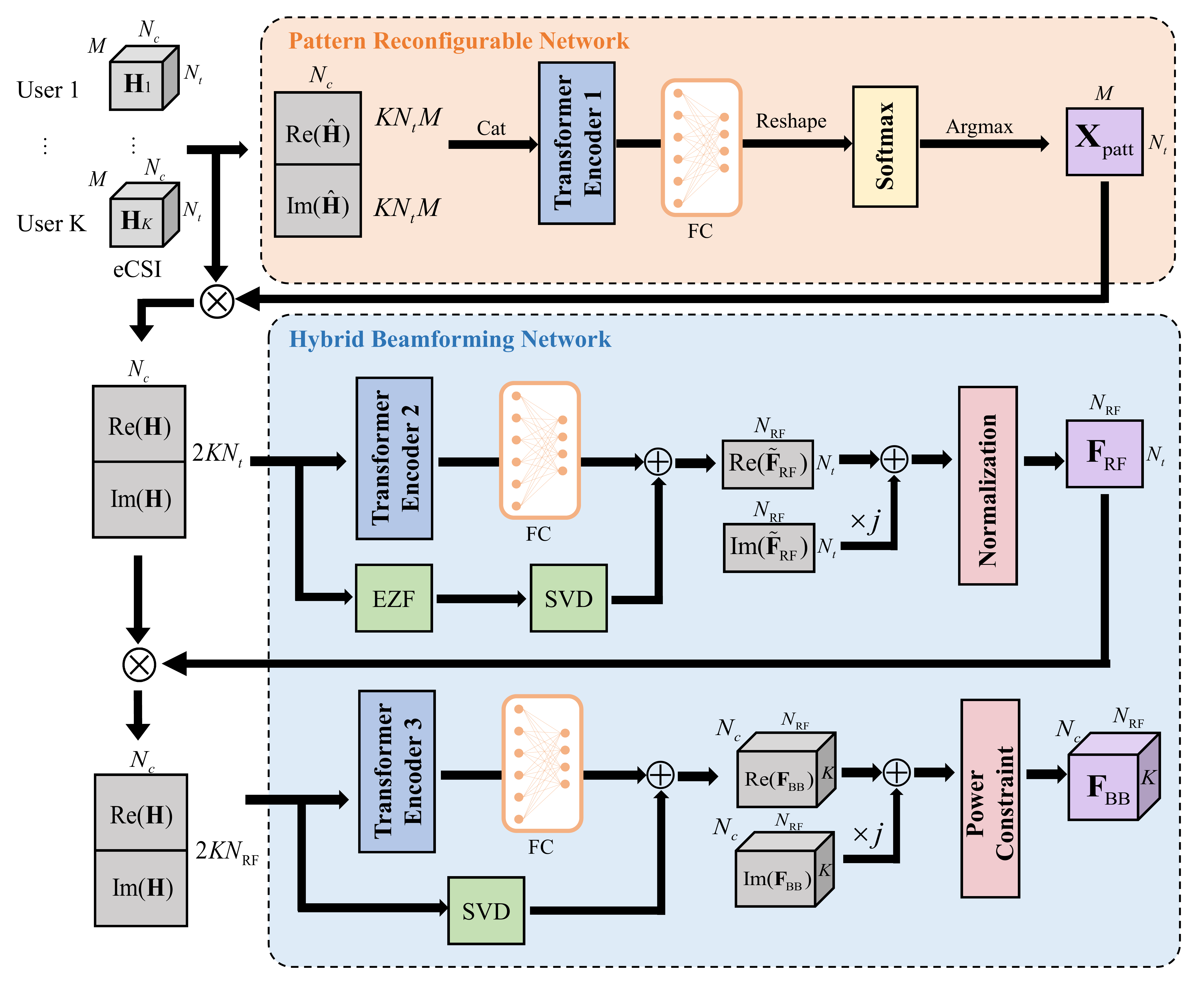}}
\vspace{-3mm}
\caption{Proposed Pattern Reconfigurable Hybrid Beamforming Network.}
\label{fig_prhbfnet}
\end{figure}

Solving the highly coupled MINLP problem in \eqref{eq:opt_prob} under the strict real-time constraints of HAPS platforms is computationally prohibitive for traditional algorithms. To address this issue, we propose an end-to-end deep learning architecture, termed PR-HBFNet, which is illustrated in Fig.~\ref{fig_prhbfnet}. The proposed PR-HBFNet is a cascaded neural network consisting of two primary modules: the Pattern Reconfigurable Network (PRN) and the Hybrid Beamforming Network (HBN). By leveraging the self-attention mechanism of Transformers and integrating model-driven residual learning, the framework significantly reduces computational complexity.

\subsection{Pattern Reconfigurable Network}
The PRN is designed to extract spatial-electromagnetic channel features and subsequently determine the radiation pattern of each antenna element. The input to the PRN is the electromagnetic channel state information tensor $\mathbf{\hat{H}}$. For $K$ users, $N_c$ subcarriers, $N_t$ antennas, and $M$ candidate patterns, the tensor is first separated into real and imaginary parts and then rearranged into a real-valued feature sequence. This sequence is fed into \textbf{Transformer Encoder 1}, whose self-attention mechanism captures the long-range dependencies among antennas, users, and candidate radiation patterns \cite{vaswani2017attention}. 

The extracted features are passed through a fully connected layer and reshaped into an $N_t \times M$ score matrix. A Softmax activation is applied along the pattern dimension to generate the probability of each candidate mode for every antenna. Finally, an Argmax operation selects the most likely radiation pattern and outputs the pattern selection vector $\mathbf{X}_{\text{patt}} \in \mathcal{M}^{N_t}$, which corresponds to the pattern variable $\mathbf{c}$ in the system model. Since Argmax is non-differentiable, a Straight-Through Estimator (STE) \cite{bengio2013estimating} is employed during training: the forward pass uses the discrete output, while the backward pass approximates the gradient using the Softmax result. 

\subsection{Hybrid Beamforming Network}
Based on the selected pattern vector $\mathbf{X}_{\text{patt}}$, we extract from $\mathbf{\hat{H}}$ the channel coefficients associated with the active radiation mode of each antenna. This produces the equivalent channel set $\{\mathbf{H}_g\}_{g=1}^{N_c}$ used by the beamforming network. The HBN is then tasked with computing the analog precoder $\mathbf{F}_{\text{RF}}$ and the digital precoders $\{\mathbf{F}_{\text{BB},g}\}_{g=1}^{N_c}$.

To improve convergence and preserve the physical structure of the beamforming problem, the HBN adopts a \textbf{model-driven residual learning} strategy \cite{ying2024reconfigurable, hu2021iterative}. 

\textbf{(1) Analog Beamforming Branch:} The equivalent channel is separated into real and imaginary parts and passed to \textbf{Transformer Encoder 2}. In parallel, a deterministic baseline based on subarray-wise SVD is used to provide an initial analog beamformer for the sub-connected architecture. The Transformer branch learns a residual correction over this initialization. The resulting analog beamformer is then projected onto the feasible sub-connected set so that the final $\mathbf{F}_{\text{RF}}$ remains block diagonal and each non-zero entry satisfies the constant-modulus constraint required by the passive phase shifters. 

\textbf{(2) Digital Beamforming Branch:} After obtaining $\mathbf{F}_{\text{RF}}$, the effective low-dimensional baseband channel on each subcarrier is formed as $\mathbf{H}_g \mathbf{F}_{\text{RF}}$. This effective channel is fed into \textbf{Transformer Encoder 3}. Similar to the analog branch, an SVD-based baseline is used to generate an initial digital precoder, while the Transformer predicts a residual refinement to better suppress multi-user interference. The final digital precoders are then jointly rescaled over all subcarriers so that the total transmit-power constraint in \eqref{eq:opt_prob} is satisfied. In this way, the normalization is consistent with the average power budget across the full OFDM block rather than being applied independently to each subcarrier. 

The entire PR-HBFNet is trained end-to-end in an unsupervised manner by directly maximizing the system spectral efficiency. The loss function is defined as the negative average sum spectral efficiency,
\begin{equation}
\mathcal{L} = -\frac{1}{N_c} \sum_{g=1}^{N_c} \sum_{u=1}^{K} \log_2(1 + \gamma_{u,g}),
\end{equation}
which eliminates the need for labeled optimal solutions from conventional solvers during training. 

By executing the pattern selection and hybrid beamforming sequentially, the proposed PR-HBFNet avoids the high latency of iterative search while preserving the key hardware constraints of RPA-assisted sub-connected HBF. This makes the framework suitable for low-latency beamforming in practical HAPS systems.

\section{Simulation Results}
In this section, we evaluate the SE and complexity performance of the proposed PR-HBFNet.
\subsection{Simulation Settings}
\begin{figure}[tbp]
\centerline{\includegraphics[width=0.5\textwidth]{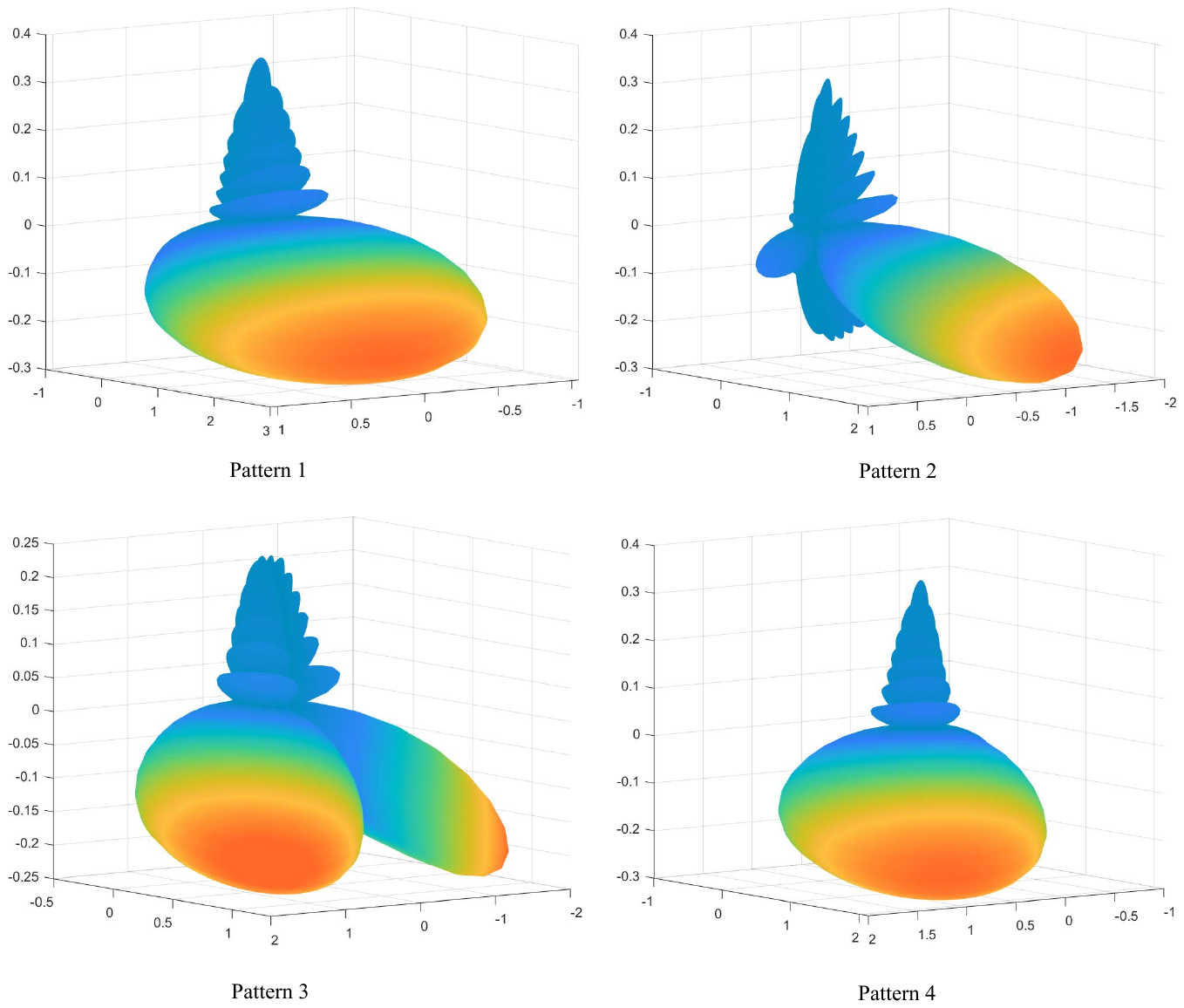}}\vspace{-3mm}
\caption{Reconfigurable radiation patterns.}
\label{fig_pattern}
\end{figure}

In the simulations, we adopt the 3D wideband channel model for HAPS environments as described in Section II. For realistic channel realization, we utilize the QuaDRiGa software package \cite{jaeckel2021quadriga}. The HAPS base station is equipped with a sub-connected array featuring $N_t = 32$ reconfigurable antenna elements. The system architecture incorporates $N_{\text{RF}} = 8$ radio-frequency chains. The communication waveform is based on an OFDM system with $N_c = 60$ subcarriers and the total signal bandwidth is $B_w = 7.2$ MHz. For the performance evaluation, the noise power spectral density is set to $-174$ dBm/Hz. The total transmit power $P_t$ is varied during simulations to evaluate performance across different signal-to-noise ratio regimes. As depicted in Fig.~\ref{fig_pattern}, we consider a pattern codebook of size $M=4$. The candidate modes are categorized by their peak gain azimuth angles ($\phi$): Pattern 1 at $0^\circ$, Pattern 2 at $30^\circ$, Pattern 3 at $\pm 56^\circ$, and Pattern 4 at $-30^\circ$. While traditional massive MIMO relies on a fixed Pattern 1 direction, the reconfigurable elements provide diverse radiation modes that offer improved spatial adaptability compared to conventional fixed-pattern arrays.

The generated channel dataset is partitioned into training, validation, and testing subsets, containing 102,400, 10,240, and 10,240 samples, respectively. The proposed PR-HBFNet is implemented on a system with an NVIDIA GeForce GTX 2080Ti GPU. Training is conducted with a batch size of 512 over 50 epochs, using the Adam optimizer for parameter updates. Additionally, a learning rate schedule with a warm-up strategy is applied to improve convergence efficiency and training robustness. 

\subsection{Performance Comparison}

In this subsection, we evaluate the sum SE performance of the proposed PR-HBFNet. To ensure a fair comparison, the benchmark denoted as \emph{Greedy+HBFNet} uses greedy pattern selection while retaining the same residual hybrid beamforming module as the proposed method. We further compare the proposed network against random pattern selection and traditional massive MIMO baselines using fixed radiation patterns. 
\begin{figure}[htbp]
\centerline{\includegraphics[width=0.4\textwidth]{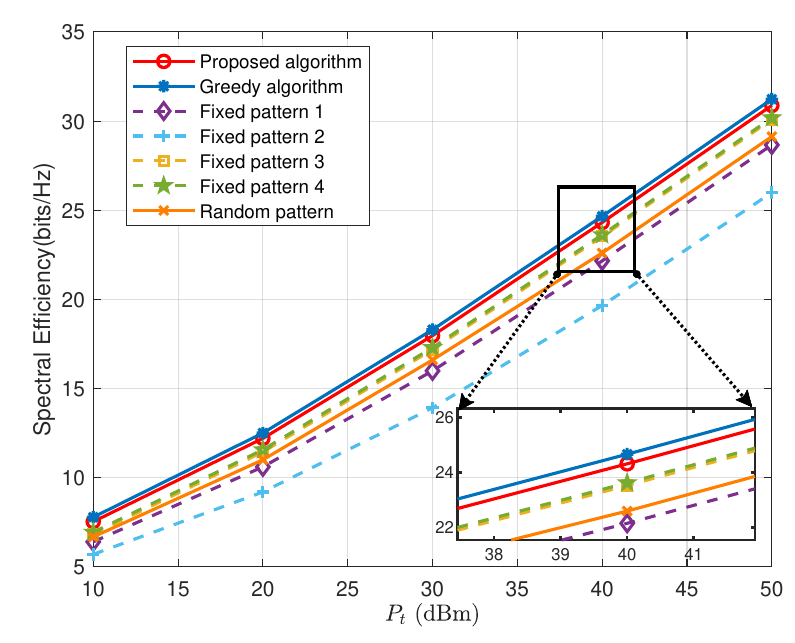}}
\caption{SE performance comparison versus $P_t$.}
\label{fig_se_pt}
\end{figure}

\begin{figure}[htbp]
\centerline{\includegraphics[width=0.4\textwidth]{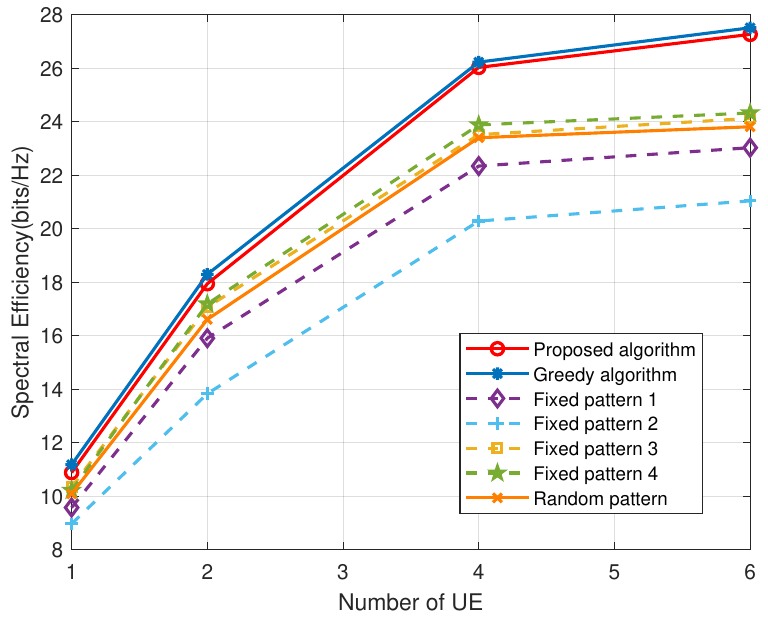}}
\caption{SE performance comparison versus number of UEs.}
\label{fig_se_K}
\end{figure}

As shown in Fig.~\ref{fig_se_pt}, all schemes employing reconfigurable antennas consistently outperform those with fixed radiation patterns, because fixed-pattern antennas lack the additional spatial flexibility provided by pattern adaptation. Notably, the proposed PR-HBFNet achieves a performance very close to the Greedy+HBFNet benchmark. For $P_t=50$ dBm, PR-HBFNet achieves an SE of approximately 30.87 bits/s/Hz, which is within 1.1\% of the benchmark value of 31.23 bits/s/Hz.

Fig.~\ref{fig_se_K} depicts SE performance versus the number of users $K$. As observed, both the proposed PR-HBFNet and the Greedy+HBFNet benchmark sustain a notable performance advantage over all fixed-pattern schemes. With the increase in $K$, multi-user interference intensifies, and the performance gap between reconfigurable antenna-based systems and traditional massive MIMO becomes more pronounced. This verifies that the proposed framework can dynamically reconfigure antenna radiation patterns and beamforming weights according to the channel geometry of different users.
\begin{table}[htbp]
\centering
\caption{Complexity Comparison}
\scriptsize
\setlength{\tabcolsep}{4pt}
\begin{tabular}{c|ccc|ccc}
\hline
$K$ & \multicolumn{3}{c|}{Proposed PR-HBFNet} & \multicolumn{3}{c}{Greedy+HBFNet} \\
\cline{2-7}
& Params & FLOPs & Time (ms) & Params & FLOPs & Time (ms) \\
\hline
1 & 9.610M & 963.25M & 26.83 & 6.093M & 7.79G & 4258.86 \\
2 & 12.847M & 1.01G & 29.05 & 9.264M & 15.61G & 5012.86 \\
4 & 25.612M & 1.12G & 33.33 & 21.898M & 31.35G & 6587.12 \\
6 & 46.766M & 1.25G & 38.30 & 42.921M & 47.23G & 8288.08 \\
\hline
\end{tabular}
\label{tab:complexity}
\end{table}
Table~\ref{tab:complexity} compares the computational complexity of the proposed PR-HBFNet and the Greedy+HBFNet benchmark for different numbers of scheduled users $K$. The proposed PR-HBFNet outperforms the benchmark significantly in terms of FLOPs and runtime: its runtime stays at the millisecond level (26.83--38.30 ms) with FLOPs ranging from 963.25 M to 1.25 G, while the benchmark requires 4258.86--8288.08 ms and 7.79 G--47.23 G FLOPs. The increase in trainable parameters of the proposed PR-HBFNet remains moderate compared with the substantial gain in computational efficiency, which supports its suitability for low-latency HAPS communication systems.

These results verify that the proposed low-complexity network can effectively approach the performance of iterative or greedy pattern-selection schemes while remaining superior to conventional fixed-pattern massive MIMO.

\section{Conclusion}
In this paper, we proposed a Transformer-based PR-HBFNet for pattern reconfigurable antenna-equipped massive MIMO in HAPS communications. The proposed network jointly optimizes radiation pattern selection and hybrid beamforming through a PRN for spatial-electromagnetic feature extraction and pattern determination, and an HBN with model-driven residual learning for analog and digital precoding. By explicitly incorporating the sub-connected analog architecture and the total transmit-power constraint into the learning framework, the proposed method remains consistent with the underlying hardware model. Simulations demonstrated that PR-HBFNet achieves SE within 1.1\% of the Greedy+HBFNet benchmark while reducing inference latency from seconds to milliseconds, confirming its potential for real-time HAPS systems.

\end{document}